\documentclass[12pt,preprint,epsfig]{aastex}

\begin{document}
\title{Hyperbolic Hamiltonian equations for general relativity}
\author{Maurice H.P.M. van Putten}
\affil{Le Studium IAS, 45071 Orl\'eans Cedex 2, Universit\'e d'Orleans, France}
\begin{abstract}
The 3+1 Hamiltonian formulation in the gauge $D_tN=-K$ on the lapse function fixes 
the direction of time associated with the trace $K$ of the extrinsic curvature tensor. 
The Hamiltonian equations hereby become hyperbolic. We study this new system for 
black hole spacetimes that are asymptotically quiescent, which introduces analyticity 
properties that can be exploited for numerical calculations by compactification in 
spherical coordinates with complex radius following a M\"obius transformation. 
Conformal flat initial data of two black holes are hereby invariant, and correspond 
to a turn point in a pendulum, up for a pair of separated black holes and down for a
single black hole. Here, Newton's law appears in the relaxation of $l=2$ deformations 
of semi-infinite poloidal surface elements, defined by the moment of inertia of the 
binary. 
\end{abstract}

\section{Introduction}

The calculation of gravitational radiation produced in the merger of two black holes is of 
considerable theoretical interest. It may also provide templates of wave-forms for analysis 
of LIGO-Virgo data. To be of practical interest to matched filtering, the calculational 
methods will have to recover many wave-periods over an extended parameter range, such as in 
the problem of binary black hole coalescence with a range of black hole masses and spins. 
It poses the challenge of efficient and stable phase-accurate numerical methods. It has been
appreciated that this requires an inherently stable formulation of the dynamical evolution
of spacetime. Spectral methods provide the most efficient representation of functions
that are periodic and everywhere analytic, and preserve accurate phase information in hyperbolic 
evolution. 

Here, we consider the 3+1 Hamiltonian equations for the dynamical behavior of spacetime \citep{arn62}. 
They are attractive, in describing the evolution in terms of quantities that have unambiguous geometric 
interpretations. However, for arbitrary choices of gauge, they are known to give rise
to computational instabilities, which have thusfar prevented their application to large-scale
numerical relativity.

Here, we propose a new gauge for time-evolution, which describes a correlation between the
lapse function and the extrinsic curvature of the foliation of three-surfaces in spacetime.
Linearized stability analysis shows that the full system of Hamiltonian equations now becomes
hyperbolic with respect to arbitrary perturbations of the three-metric. Next, we focus on
black hole spacetimes which are asymptotically quiescent \citep{van06}. This asymptotic property, 
if present in the initial data, is preserved for all time by causality in the equations of 
general relativity. It can be exploited by compactification in spherical coordinates with complex 
radius, $(z,\theta,\phi)$ with complex radius $z$, $-\infty + is < z < \infty + is$, where $E/s$ 
denotes a dimensional continuation parameter for a spacetime with total mass energy $E$. Here, 
$s$ can be chosen sufficiently large, i.e., at least on the order of $E$, to avoid singularities 
associated with black holes, even those that may carry spin. In this approach, we focus on the 
{\em outer} expansion of the Green's function. It makes possible a three-dimensional spectral 
representations after application of a M\"obius transformation (which preserves conformally flat 
initial data).

The combination of hyperbolic Hamiltonian equations and spectral representations offers a novel
starting point for studying the dynamical evolution of black hole spacetimes. For illustrative
purposes, we consider the conformally flat initial data of Schwarzschild black holes, as in
the problem of a head-on collision of two black holes. Here, conformally flat data are special, 
in representing turnings points corresponding to the up or down position of a pendulum. 
The proposed lapse function is natural choice about such turning points. To illustrate compactification
by a M\"obius transformation, we explicitly identify Newton's law in terms of the evolution of the 
quadrupole deformations in area of poloidal flux surfaces, given by the moment of inertia $I$ of 
spacetime. 

\section{Asymptotic wave motion}

At large distances from the source region, the two wave-modes in the gravitational field 
appear as planar waves in the three-metric $h_{ij}$, commonly expressed in the 
transverse traceless gauge as
\begin{eqnarray}
h_{ij}=\left(\begin{array}{ccc}
1 & 0 & 0\\
0 & 1+b & c\\
0 & c   & 1-b
\end{array}\right), ~~b=B\sin(kz),~~c=C\sin(kz)
\label{EQN_A1}
\end{eqnarray}
$B,C$ denote the amplitude of $+$ and $\times$ plane-waves along the $z-$direction with wave-number 
$k$. The three-dimensional Ricci tensor associated with $h_{ij}$ hereby assumes the following
structure. For the $+$ waves with $(B,C)=(B,0)$, we have
\begin{eqnarray}
R_{ij}^+=\frac{Bk^2}{2\Delta}\left(\begin{array}{ccc}
* & 0 & 0\\
0 & -\sin kz - B\cos^2kz + B^2\sin^3 kz & 0\\
0 & 0 & \sin kz - B\cos^2kz - B^2\sin^3 kz
\end{array}\right),
\label{EQN_A2}
\end{eqnarray}
where $b=B\sin kz,$ $c=0$ and $\Delta = 1- B^2\sin^2 kz$, and for the $\times$ waves with 
$(B,C)=(0,C)$, $b=0$, $c=C\sin kz$, we find, similarly,
\begin{eqnarray}
R_{ij}^\times=\frac{Ck^2}{2\Delta}\left(\begin{array}{ccc}
* & 0 & 0\\
0 & -C\cos^2 kz & - \Delta \sin kz\\
0 & -\Delta \sin kz  &  - C\cos^2kz
\end{array}\right),
\label{EQN_A3}
\end{eqnarray}
where $\Delta = 1-C^2\sin^2 kz$. Hence, we have the following exact algebraic identities:
\begin{eqnarray}
R_{22}^+-R_{33}^+\equiv -Bk^2\sin(kz),~~R_{23}^+\equiv0,\\
R_{33}^\times-R_{22}^\times \equiv 0,~~
R_{23}^\times \equiv -\frac{C}{2}k^2\sin(kz),
\label{EQN_ID}
\end{eqnarray}
where the numerical indices refer to normalized directions in the coordinate directions $(r,\theta,\phi)$.

The identies (\ref{EQN_ID}) show that the two polarization modes satisfy traveling wave solutions
\begin{eqnarray}
b(t,r)=B\sin\eta,~~c(t,r)=C\sin\eta, ~~\eta = k(t-r)
\end{eqnarray} 
in the large-distance approximation (neglecting $1/r$ curvature terms, in this section) 
with vanishing lapse functions $\beta_i\equiv0$ and constant lapse function $N\equiv1$, satisfying
\begin{eqnarray}
\partial_t^2 \left(h_{22} - h_{33}\right) = -2\partial_t(K_{33}-K_{22})= 2(R_{22}-R_{33}),~~
\partial_t^2 h_{23} = -2\partial_t K_{23}= 2 R_{23},
\end{eqnarray}
where $b(t,r)=\frac{1}{2}\left(h_{22}-h_{11}\right),~~c(t,r)=h_{12},$ using the numerical indices
to denote the normalized directions in $(r,\theta,\phi$). The angular directions ``22-33" and ``23" 
therefore represent the {\em hyperbolic directions}, which contain all wave-motion.

\section{General 3+1 decomposition of the metric}

We consider the 3+1 decomposition of the line-element in terms of $h_{ij}$ \citep{tho86}
\begin{eqnarray}
ds^2 =-N^2dt^2+h_{ij}\left(dx^i+\beta^idt)(dx^j+\beta_jdt\right).
\end{eqnarray}
The lapse and shift functions can be seen to define the motion of observers with 
velocity four-vector $u^i=(1,0,0,0)$, for example, by considering their acceleration
$a^i=\frac{d}{dt}u^i= - \Gamma_{00}^i 
   = -\frac{1}{2}g^{ii}\left(2\dot{g}_{i0}-g_{00,i}\right) 
   = -h^{ii}\left(\dot{\beta}_i-N\partial_iN+\beta^jD_i\beta_j\right)
   = \Phi^{-4}N\partial_iN,$
which is determined entirely by the redshift factor $N$ when $\beta_i=\dot{\beta}_i=0$. In
general, the lapse and shift functions appear in the conservation of energy and momentum,
of test particles moving along geodesics. We need not impose the time-symmetric gauge condition 
with $\dot{\beta}_i=0$. The three-metric $h_{ij}$ satisfies the Hamiltonian evolution equations 
\begin{eqnarray}
\partial_th_{ij} = D_i\beta_j + D_j\beta_i - 2N K_{ij},~~
N^{-1}D_t K_{ij} +2K_i^mK_{jm}-KK_{ij} = -W_{ij},
\label{EQN_BM2}
\end{eqnarray}
where $D_tK_{ij} = \partial_tK_{ij} - \beta^mD_mK_{ij} - K_{im}D_j\beta^m - K_{jm}D_j\beta^m$
and $W_{ij}$ denotes the gauged three-tensor $W_{ij}=-R_{ij}+ N^{-1}D_iD_jN.$ 

We propose the time-evolution accoring to
\begin{eqnarray}
K:\left\{
\begin{array}{lll}
D_tN & = &-K,\\
D_th_{ij} & =& - 2K_{ij}N,\\
D_t K_{ij} &=& -D_{ij}N+{\cal K}_{ij}N,
\end{array}\right.
\label{EQN_K}
\end{eqnarray}
where ${D}_{ij}=D_iD_j-R_{ij}$ and ${\cal K}=KK_{ij} -2K_i^mK_{jm}$.

The gauge condition $D_tN=-K$ is {\em curvature-driven evolution}. 
It is different from the product of curvature and lapse function in
the harmonic slicing condition $\partial_t N = -N^2 K$ in Eqs. (69)-(77) of \cite{abr97}; 
see also \cite{bro08} for a recent review.

In the asymptotically flat region about $h_{ij}=\delta_{ij}$ and $N=1$, we have
\begin{eqnarray}
\partial_t N = -K,~~\partial_t^2 h_{ij} = -2R_{ij}+2D_iD_jN,~~\partial_t^2 K = \Delta K.
\end{eqnarray}
We recall that \citep{wal84}
\begin{eqnarray}
R_{ij} = -\frac{1}{2} \Delta \delta h_{ij} + 
         +\frac{1}{2} \partial_i\partial^e\bar{\delta h}_{ej}
         +\frac{1}{2}\partial_j\partial^e\bar{\delta h}_{ei}
\end{eqnarray}
where $\bar{h}_{ij} = \delta h_{ij} -\frac{1}{2}\delta_{ij} \delta g$, where 
$\delta h=h^{ij}\delta h_{ij}$ refers to the trace of the metric perturbations.
For harmonic perturbations of the for 
$\delta h_{ij} \sim \hat{h}_{ij}e^{-i\omega t}e^{ik_ix^i}$,
we have, with conservation of momentum, $D^iK_{ij} = D_jK$,
$-i\omega N=- \hat{K},~~\hat{\delta h}_{ij} = - 2i\omega^{-1}\hat{K}_{ij},~~
k^i\hat{K}_{ij} = k_j\hat{K}$, so that
\begin{eqnarray}
\partial_i\partial^e\bar{h}_{ej} \rightarrow 
 k_ik^e\hat{h}_{ej} - \frac{1}{2}k_ik_j\hat{{\delta h}}
= i\omega^{-1} (-2 k_ik^e \hat{K}_{ej}+k_ik_j\hat{K})=-i\omega^{-1}k_ik_j\hat{K}
\end{eqnarray}
and hence
\begin{eqnarray}
\hat{R}_{ij} - \partial_i\partial_j N 
= \frac{1}{2} k^2 \hat{h}_{ij} - i\omega^{-1}k_ik_j\hat{K} + i\omega^{-1}k_ik_j\hat{K}
= \frac{1}{2} k^2 \hat{h}_{ij}
\end{eqnarray}
We conclude that $\partial_th_{ij}=-2R_{ij}+2D_iD_jN$ gives rise to the dispersion
relation
\begin{eqnarray}
\omega^2=k^2
\label{EQN_D}
\end{eqnarray}
for arbitrary, small amplitude metric perturbations.
We conclude that the system $K$ in (\ref{EQN_K}) is asymptotically stable by virtue
of the gauge choice $D_tN=-K$.
This contrasts with the more common derivation of (\ref{EQN_D}) in the so-called
transverse traceless gauge, or harmonic coordinates--neither of these two coordinate
conditions are used here.

\section{Conformal flat initial data}

The conformal decomposition of the Ricci tensor for $h_{ij}=\phi^{2m}g_{ij}$ satisfies \citep{wal84}
$^{(n)}R_{ij}(h)=^{(n)}{R}_{ij}(g) - m\phi^{-1}[(n-2){D}_i{D}_j\phi + {g}_{ij} {\Delta} \phi]
+m\phi^{-2}[(1+m)(n-2){D}_i\phi{D}_j\phi + (1-m*(n-2)){g}_{ij} {D}^p\phi {D}_p\phi].$
For $n=3$ and $m=2$, it reduces to
$^{(3)}R_{ij}=^{(3)}{R}_{ij}(g) - 2\phi^{-1}[{D}_i{D}_j\phi + {g}_{ij} {\Delta} \phi]
       + 2\phi^{-2}[3 {D}_i\phi{D}_j\phi - {g}_{ij} {D}^p\phi {D}_p\phi],$
giving the familiar result 
\begin{eqnarray}
R_h=\phi^{-4}\left[R_g- 8\phi^{-1}{\Delta}_g\phi\right].
\label{EQN_CID}
\end{eqnarray}

\begin{figure}
\centerline{\includegraphics[width=80mm,height=80mm]{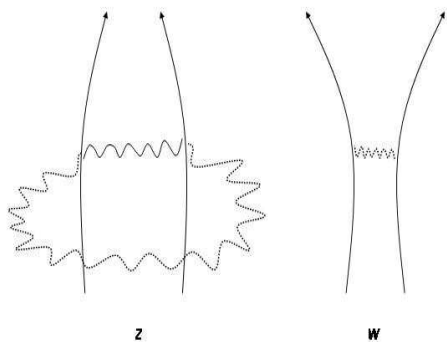}}
\caption{{\small
The interaction of two Schwarzschild black holes is attractive as 
seen in the space of the complex radial coordinate $z$ (solid curved line) 
or equivalently repulsive, following closure over infinity in the domain of convergence 
of the outer expansion of the Green's function, i.e., in the space of the transformed 
variable $w=1/z$ (dotted curved lines). For the latter, Newton's law represents
the leading order evolution of the area of the poloidal surface elements $|z|>p$,
governed by the moment of inertia associated with the separation $p$ between the 
two black holes.}}
\end{figure}

Time-symmetric initial data around one or multiple non-rotating black holes can be
conveniently described in a conformally flat representation \citep{lin63,bril64,jan03} on
the basis of (\ref{EQN_CID}).
It gives rise to solutions for vacuum spacetimes in terms of the Green's function $G(x^i,p^i)=G_p$ 
of flat spacetime, where $p^i$ denotes the position of a point source.

Spacetime of a single black hole is given by the conformal factor
\begin{eqnarray}
\Phi = 1 + 2\pi m G_p = 1 + \frac{E}{2z} + \frac{mp}{2z^2} P_1(x) + \frac{I}{2z^3}P_2(x)+\cdots
\end{eqnarray}
where $E=m$, $I=mp^2$ and the expansion refers to the {\em outer expansion} of $G_0$ in complex 
spherical coordinates $(z,x,\phi)$, $x=\cos\theta$, by exploiting analyticity at infinity, i.e.,
asymptotic quiescence. The same construction applies to two black holes,
\begin{eqnarray}
\Phi = 1 + 2\pi M G_q + 2\pi m G_p = 1 + \frac{E}{2z} + \frac{Mq+mp}{2z^2}P_1(x) + \frac{I}{2z^3}P_2(x)+\cdots
\label{EQN_PHI}
\end{eqnarray}
where $E=M+m$ and $I=Mq^2+mp^2$. Without loss of generality, we may choose $Mq+mp=0$.
According to the Hamiltonian evolution equations, the associated shift function, for a static
foliation of spacetime, i.e., $D_tK_{ij}=0$, satisfies 
\begin{eqnarray}
\Delta_h N + 3NR=0.
\end{eqnarray}
For the conformally flat data at hand, it reduces to $(\Phi^2\partial_iN)_i=0$, or
$N=(2-\Phi)/\Phi$.

The leading-order structure of spacetime can be studied by considering poloidal surface elements 
$dA=h_{\theta\theta}^{1/2}h_{zz}^{1/2}d\theta dz$ in the outer region $|z|>p$. A measure for the 
$l=2$ deformations in surface elements is
\begin{eqnarray}
A_2(z,t)=\int_z^\infty \int_0^\pi dA= 
    \int_z^\infty \int_0^\pi \Phi^4 zd\theta dz +  \cdots
\end{eqnarray}
in $z>p$, where $\cdots$ refers to time-independent divergences. It gives a geometric 
equivalence about $w=0$ $(w=1/z)$ to the moment of intertia.

\section{Newton's law in the initial quadrupole evolution}

The gauged tensor $W_{ij}=-R_{ij}+N^{-1}D_iD_jN$ has a regular Taylor series expansion about 
infinity, for the given asymptotically quiescent initial data.
For the equal mass, two-hole solution with total mass-energy $E=2M$,
moment of intertia $I=2Mp^2$, and symmetric position about the origin,
the conformal factor 
\begin{eqnarray}
\Phi=1+\frac{E}{2z}+\frac{I}{2z^3}P_2(x)+\cdots
\end{eqnarray}
we find the large-$z$ asymptotics 
\begin{eqnarray}
W^*_{ij}= -\frac{EI}{2z^6}\left(\begin{array}{ccc}
2P_2 & -3zP_1 & 0\\-3zP_1& -z^{2}[2P_2-P_0]/(1-x^2)& 0\\0& 0 & -z^{2}P_0(1-x^2)
\end{array}
\right),
\label{EQN_WII}
\end{eqnarray}
satisfying $W^*=0$ with
\begin{eqnarray}
W^*:W^*=E^2I^2\left(\frac{3}{2}z^{-12}-9Ez^{-13}+\left(\frac{117}{4}E^2-9E^{-1}Ix^2\right)z^{-14}\right)
+O(z^{-15})
\end{eqnarray}

The time-evolution of the $l=2$ deformation of poloidal surface elements is given by
\begin{eqnarray}
\ddot{A}_2(z,t)=\int_0^{\pi}\int_{z}^\infty 
  \partial_t^2 \mbox{det}(h_{IJ})^{1/2} d\theta dz
\end{eqnarray}
where $I,J=z,\theta$, where
$\mbox{det}(h_{IJ})=h_{\theta\theta} h_{zz}-(h_{z\theta})^2.$
For the time-symmetric initial data, the quadratic 
off-diagonal terms make no initial contribution, leaving
\begin{eqnarray}
\ddot{A}_2(z,t)=\int_0^{\pi}\int_{z}^\infty 
  \partial_t^2 (h_{\theta\theta}^{1/2}h_{zz}^{1/2})d\theta dz =
  \frac{1}{2}\int_0^{\pi}\int_{z}^\infty
  (h_{\theta\theta}^{-1/2} h_{zz}^{1/2} \ddot{h}_{\theta\theta}+
   h_{zz}^{-1/2}h_{\theta\theta}^{1/2}\ddot{h}_{zz})d\theta dz.
\end{eqnarray}
At $t=0$, the integrand reduces in the gauge $K=0$ with the additional
time-symmetric gauge condition $\dot{\beta}_i=0$ at $t=0$ (a Newtonian
gauge choice, as in a turning point) to
\begin{eqnarray}
  \frac{1}{2}\left(\ddot{h}_{zz}+z^{-2}\ddot{h}_{\theta\theta}\right)z
=-\frac{1}{2z^2\sin\theta}\ddot{h}_{\phi\phi}
= z^{-2}\sin^{-2}\theta \dot{K}_{\phi\phi} = -\frac{EI}{2z^4}+O(z^{-5}).
\end{eqnarray}
For completely time-symmetric initial data, we thus recover Newton's law for
two equal mass black holes of mass $M$ and separation $2p$ in 
$\ddot{A}_2(z,t)=-\frac{1}{2}EIz^{-4}+O(z^{-5})$, we find
\begin{eqnarray}
\ddot{p}=-\frac{M}{4p^2}
\label{EQN_NEW}
\end{eqnarray}
for initial evolution of $A_2$ at the maximal extent $z=p$ (the lower limit of $z$)
of the poloidal surface elements (in the outer expansion considered).
 
Thus, for a fixed $|z|>p$, it takes $t=O(z^3)$ to relax $Iz^{-3}$ by the leading-order
$z^{-6}$ forcing terms in (\ref{EQN_WII}), when applied to the asymptotic equations.
The asymptotics $Iz^{-3}$ for large $z$ hereby persists for 
all finite time. This observation is consistent with the a priori notion of preserving 
asymptotic quiescence. Here, Newton's law (\ref{EQN_NEW}) associated with the deformations
of the poloidal surface elements $|z|>p$ represent the underlying
non-local interaction.

\section{Conclusions}

We have shown that the 3+1 Hamiltonian equations become hyperbolic, when the evolution
of the lapse function is driven by the extrinsic curvature tensor. The resulting
hyperbolic system forms a starting point for three-dimensional spectral methods in
numerical relativity, after compactification by a M\"obius transformation for spacetimes
that satisfy asymptotic quiescence, wherein Newton's law arises in the dynamics of the
quadrupole deformations of poloidal surface elements. Future work will focus on some
numerical examples.
 
{\bf Acknowledgment.} The author gratefully acknowledges stimulating discussions with
A. Spallicci, M. Volkov, G. Barles, and members of the F\'ed\'eration Denis Poisson.
This work is supported, in part, by Le Studium IAS of the Universit\'e d'Orl\'eans.


\end{document}